\documentclass[aps,pre,nobibnotes,showpacs,twocolumn]{revtex4-1}
\usepackage{graphicx}
\usepackage{bm}
\usepackage{amsmath}
\usepackage{amssymb}
\usepackage{euscript}
\usepackage{verbatim}
\usepackage{fancyhdr}
\usepackage{wrapfig}
\usepackage{setspace}
\usepackage{xcolor}
\usepackage{amsfonts}
\usepackage{subfigure}

\newcommand{\be}{\begin{equation}}
\newcommand{\ee}{\end{equation}}
\newcommand{\bea}{\begin{eqnarray}}
\newcommand{\eea}{\end{eqnarray}}

\newcommand{\modified}[1]{{#1}}

\begin{document}

\begin{titlepage}
	

	\title{Symmetry-resolved entanglement in many-body systems}

	\author{Moshe Goldstein}
	\email{Equal contribution}
	\affiliation{Raymond and Beverly Sackler School of Physics and Astronomy, Tel-Aviv University, Tel Aviv 6997801, Israel}
	\author{Eran Sela}
	\email{Equal contribution}
	\affiliation{Raymond and Beverly Sackler School of Physics and Astronomy, Tel-Aviv University, Tel Aviv 6997801, Israel}

	\begin{abstract}
		Similarly to the system Hamiltonian, a subsystem's reduced density matrix is composed of blocks characterized by symmetry quantum numbers (charge sectors).
		We present a geometric approach for extracting the contribution of individual charge sectors to the subsystem's entanglement measures within the replica trick method,
		via threading appropriate conjugate Aharonov-Bohm fluxes through a multi-sheet Riemann surface. Specializing to the case of 1+1D conformal field theory, we obtain general exact results for the entanglement entropies and spectrum, and apply them to a variety of systems, ranging from free and interacting fermions to spin and parafermion chains, and verify them numerically.
		We find that the total entanglement entropy, which scales as $\ln L$, is composed of $\sqrt{\ln L}$ contributions of individual subsystem charge sectors for interacting fermion chains, or even $\mathcal{O} (L^0)$ contributions when total spin conservation is also accounted for.
		We also explain how measurements of the contribution to the entanglement from separate charge sectors can be performed experimentally with existing techniques.
	\end{abstract}
	
	
	\maketitle
	
	\draft
	
	\vspace{2mm}
	
\end{titlepage}

\emph{Introduction--}
One cannot overestimate the importance of entanglement as a fundamental aspect of quantum mechanics~\cite{einstein1935can,schrodinger1935gegenwartige,horodecki2009quantum}. Its central measure, the entanglement entropy (EE), has proven indispensable for characterizing quantum correlations and phase transitions in many-body quantum systems in condensed matter and high energy physics~\cite{amico2008entanglement,calabrese2009entanglement,laflorencie2016quantum}.
Moreover, the performance of tensor-network algorithms for many-body systems strongly depends on the scaling properties of the EE with the subsystem size~\cite{white1992density,schollwock2005density,verstraete2008matrix}.
	
The computation of the EE often involves the replica trick, where one introduces $n$ copies of the system.  The $n$th R\'{e}nyi entropy (RE) is defined as $s_n={\mathrm{Tr}} \rho_A^n$, where $\rho_A = {\mathrm{Tr}}_B \rho$ is the reduced density matrix of subsystem $A$; the EE is then $\mathcal{S}=-{\mathrm{Tr}} \rho_A \ln \rho_A = -\lim_{n \to 1} \partial_n s_n$.
Recently, this theoretical tool became an experimental method which allowed, for the first time, to extract the RE in a bosonic cold atomic system,
by preparing a twin of the  many-body quantum ground state and using an appropriate swap operation~\cite{islam2015measuring}.  



\begin{figure} 
	\centering	
	\hspace*{-.25in}
	\includegraphics[scale=0.3]{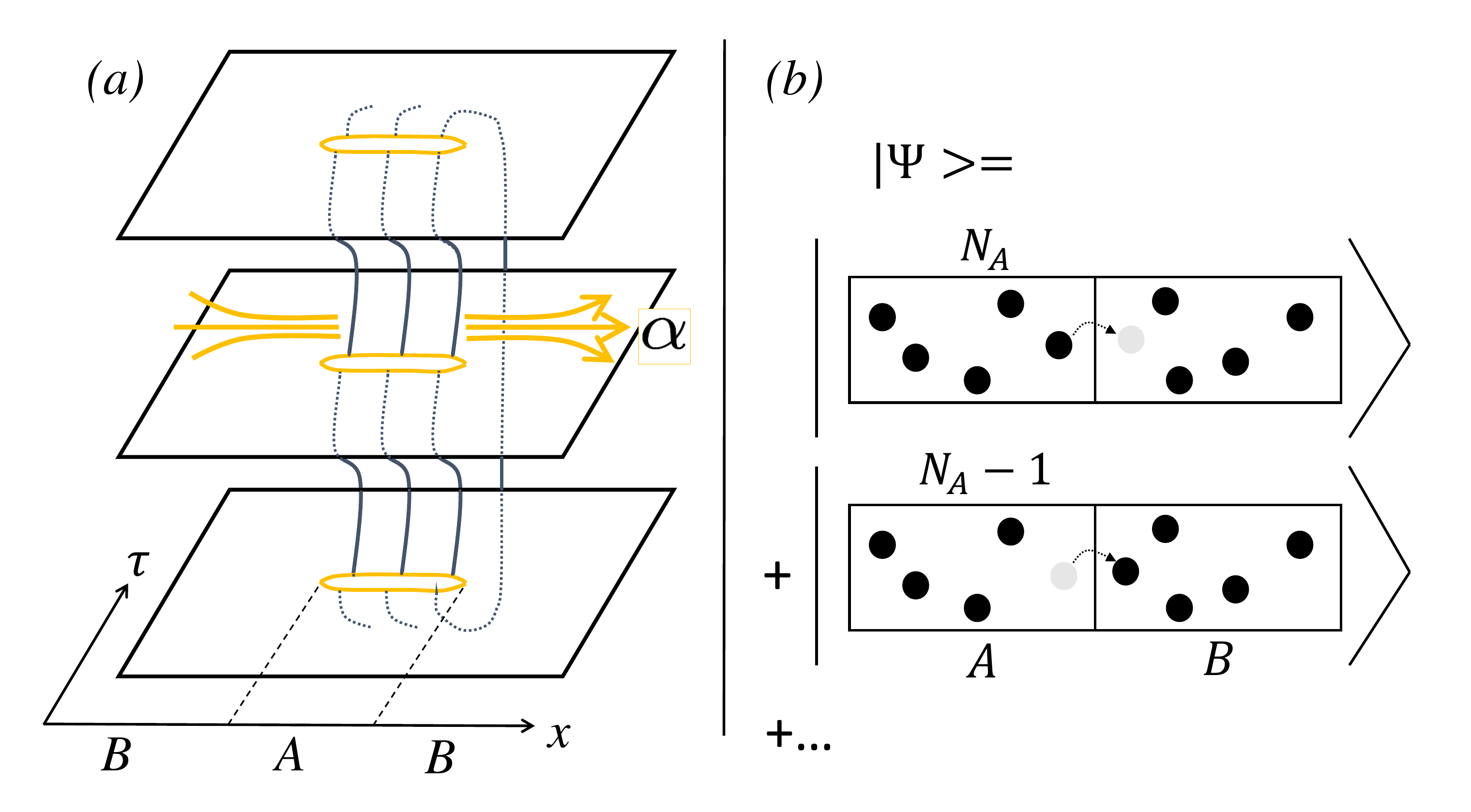}
	\caption{(Color online) (a) An example of a 3-sheet Riemann surface geometry with an inserted \modified{space-time Aharonov-Bohm} flux $\alpha$.
	(b) A generic many body wavefunction is a superposition of subsystem charge states.
	}
	\label{fig:1}
\end{figure}

Following the general path-integral approach~\cite{calabrese2009entanglement}, introducing time as a dimension, the computation of $s_n$ acquires a geometrical meaning: in the same way that the partition function of a 1D quantum system corresponds to a path integral over a cylinder with circumference given by the inverse temperature $T$, at  $T=0$ the calculation of  $s_n$ corresponds to computing the partition function on a Riemann surface geometry $\mathcal{R}_n$. For a 1D quantum system with $A$ a segment of length $L$, the Riemann geometry with $n=3$ is depicted in Fig.~\ref{fig:1}(a).

A curious question that motivated this work is: What is the physical meaning of inserting a space-time  Aharonov-Bohm flux $\alpha$ into this space, coupled, e.g., to the particles charge? When a charged particle moves from one copy to the next until it finally returns back to the initial copy, it acquires a phase $\alpha$. Thus, the total acquired phase is given by $\alpha N_A$, where $N_A$ is the total charge (number of charged particles) in region $A$ [see Fig.~1(b)]. Hence, the computation of the path integral in the presence of this flux gives the quantity
\be
\label{eq:definesn}
 s_n(\alpha) = {\mathrm{Tr}} \left( \rho_A^n e^{i \alpha \hat{N}_A} \right). 
\ee 
A related quantity (with different normalization) has recently been computed for several specific models
with holographic duals~\cite{belin2013holographic,belin2015charged,pastras2014charged}
or nontrivial topology~\cite{matsuura2016charged}, but its physical meaning has remained obscure. Here we not only provide such a meaning, but also give general results and particular examples for 1D critical systems described by conformal field theory (CFT)~\cite{yellow}, and provide recipes for its experimental measurement using the setup of~\cite{islam2015measuring}.

\emph{Main idea.---}
The theme of this work can be presented via an elementary example of a single particle located in either one of two sites, described by the wave function $|\Psi \rangle = c_1 |1 0 \rangle + c_0 |01 \rangle $.
The reduced density matrix for the first site is $\rho_A = |c_1|^2 |  1 \rangle \langle 1 | + |c_0|^2 |  0 \rangle \langle 0 |$, so the RE is $s_n = \mathrm{Tr} \rho_A^n = |c_1|^{2n} + |c_0|^{2n} $. Is it possible to separately extract each of these two terms, which are evidently associated with the region $A$ charge sectors $N_A=1$ and $0$, respectively
\footnote{Our definition of the entanglement entropy corresponds to the ``entanglement of modes'' in the terminology of Ref.~\cite{wiseman03}}?
Clearly this cannot be achieved by simply performing a charge measurement in region $A$, which would quench the entanglement.

More generally, let us assume that the density matrix of the total system $\rho$ commutes with a conserved quantity $\hat{N}$ (e.g., the total system is in a pure eigenstate of $\hat{N}$), which is a sum of contributions of the two subsystems, $\hat{N} = \hat{N}_A + \hat{N}_B$ (the nonabelian case will be discussed later on). Tracing the equation $[\hat{N}, \rho]=0$ over the degrees of freedom of subsystem B, we find that $[\hat{N}_A, \rho_A] = 0$, i.e., that $\rho_A$ is block-diagonal, with different blocks corresponding to different eigenspaces of $\hat{N}_A$ (charge sectors). Thus, the entropies are sums of contributions of the different sectors, $s_n = \sum_{N_A} s_n(N_A)$.
The individual contributions $s_n(N_A)$ were calculated numerically for a particular model~\cite{laflorencie2014}, but how could one resolve them analytically in general?
Here comes the connection with the replica trick: the symmetry resolved entropies $s_n(N_A)$ are simply the Fourier transform of the partition function on the $n$-sheet Riemann surface with a generalized \modified{Aharonov-Bohm} flux:
\bea
\label{eq:saNA}
s_n(N_A) = \int_{-\pi}^\pi \frac{d \alpha}{2 \pi} s_n(\alpha) e^{- i \alpha N_A} = {\mathrm{Tr}} \left( \rho_A^n \mathcal{P}_{N_A} \right),
\eea
where $\mathcal{P}_{N_A}$, the projector into the subspace of states of region $A$ with charge $N_A$, is the Fourier transform of $e^{i \alpha \hat{N}_A}$.


\emph{General CFT result.---}
Having defined the geometry $\mathcal{R}_n$ of an $n$-sheet Riemann surface pierced by an \modified{Aharonov-Bohm} flux $\alpha$, we now consider critical 1D systems and obtain a general exact result for the $n$th RE, which we will employ in various physical examples below. 
	
The $n$-sheet Riemann geometry pierced by a flux may be viewed as an extension of the theory into $n$ copies $\phi \to \phi_l~(l=1,...,n)$, where the fields $\phi_l$ satisfy the boundary condition $\phi_l(x,\tau=0^-) = \phi_{l+1}(x,\tau=0^+) e^{i \alpha \delta_{l,j}}$~($x \in A$), and $\phi_l(x,\tau=0^-) = \phi_{l}(x,\tau=0^+)$~($x \in B$). Here, we have chosen to insert the \modified{Aharonov-Bohm} phase in the link between copies $j$ and $j+1$. We have also made the assumption of a $U(1)$ symmetry, which will be generalized below. As suggested in~\cite{calabrese2004entanglement,cardy2008form}, in the absence of flux, one can define a local twist field $\mathcal{T}$ living at the end points of region $A$, denoted $w$ and $w^\prime$ ($w-w^\prime=L$), which generates the twisted boundary conditions with $\alpha=0$. We incorporate the additional \modified{Aharonov-Bohm} phase into the boundary condition by ``fusing'' this twist field $\mathcal{T}$ with the operator $\mathcal{V}$ generating a phase $\alpha$ for particles moving around it in sheet $j$, resulting in the composite twist field
$\mathcal{T}_\mathcal{V} = \mathcal{V}\mathcal{T}$. One may view $\mathcal{T}_\mathcal{V}(w)$ as an additional field in the $n$-copy theory $\mathcal{C}^n$, such that any correlation function on the $n$-sheet Riemann surface $\mathcal{R}_n$ with \modified{Aharonov-Bohm} flux $\alpha$ is given by 
	\be
	\label{corrfnc}
	\langle \mathcal{O}(z) \rangle_{\mathcal{R}_n,\alpha} = \frac{\langle \mathcal{O}(z)  \mathcal{T}_\mathcal{V}(w) \mathcal{T}_\mathcal{V}(w') \rangle_{\mathcal{C}^n} }{\langle  \mathcal{T}_\mathcal{V}(w) \mathcal{T}_\mathcal{V}(w') \rangle_{\mathcal{C}^n}}.
	\ee
In order to fully characterize the properties of our composite twist field $\mathcal{T}_\mathcal{V}$, we follow Ref.~\cite{calabrese2009entanglement} and uniformize the n-sheet Riemann surface into a single plane with a left over flux via a conformal transformation. Relegating the derivation to the Supplemental Material \footnote{See Supplemental Material for technical details, which includes Refs.~\cite{cardy10,alcaraz2011}.},
we find that the composite twist field has scaling dimension
\be
\label{scalingD}
\Delta_n(\alpha)=\frac{ c(n-n^{-1})}{24}+\frac{\Delta_\mathcal{V}}{n}.
\ee
Here $c$ is the central charge of the CFT, and $\Delta_\mathcal{V}$ is the scaling dimension of the operator $\mathcal{V}$ generating the generalized \modified{Aharonov-Bohm} phase.
This twist field correlator then yields our general result for the RE~\cite{calabrese2009entanglement},
\be
\label{genresult}
s_n(\alpha) \sim L^{-\frac{c}{6}(n-n^{-1})} L^{-2 \frac{\Delta_\mathcal{V}+\bar{\Delta}_\mathcal{V}}{n}},
\ee
where $\bar{\Delta}_\mathcal{V}$ is the scaling dimension of the anti-holomorphic part of $\mathcal{V}$.

\emph{U(1) charge.---}
In  this section we exemplify our general result Eq.~(\ref{genresult}) for a generic spinless fermionic chain described by a $c=1$ CFT which is equivalent to 1D massless bosons~\cite{gogolin}. Using the bosonization relation $\psi \sim e^{i \phi}$, one can implement the phase $e^{i \alpha}$ accumulated upon taking a fermion around $w$ or $w'$ in copy $j$, by inserting the vertex operator $\mathcal{V}=e^{i \frac{\alpha}{2 \pi} \phi_j}$.
For a system of \emph{interacting} fermions generically described by 
a Luttinger liquid with parameter $K$, the scaling dimension becomes
$\Delta_\mathcal{V}=			\bar{\Delta}_\mathcal{V} =\frac{1}{2}\left( \frac{\alpha}{2\pi} \right)^2K$,
such that
\footnote{\modified{In the vertex operator one may shift $\alpha$ by integer multiples of $2\pi$. Since the most relevant of these dominates, Eq.~(\ref{mainresult}) applies for $\alpha\in[-\pi,\pi]$, and should be continued periodically outside this range.
For $K=1$ Eq.~(\ref{mainresult}) can also be obtained by going to a basis of copy-decoupled fermions~\cite{belin2013holographic,PhysRevB.96.075153,cornfeld18}.}}
\be
\label{mainresult}
s_n(\alpha) = s_n(\alpha=0) L^{-\frac{2 K}{n} \left( \frac{\alpha}{2\pi} \right)^2}.
\ee
Assuming $\ln(L) \gg 1$ (with appropriate dimensionless $L$, e.g., lattice site number), the integral in Eq.~(\ref{eq:saNA}) gives for the symmetry resolved RE
\be
\label{explicitsn}
s_n(N_A) \cong s_n(\alpha=0) \sqrt{ \frac{\pi n}{2 K \ln L}} e^{-\frac{n \pi^2 \Delta N_A^2}{2 K \ln L}},
\ee
with $\Delta N_A = N_A - \langle N_A \rangle$.
For $n=1$, $s_1(\alpha)$ is the generating function of the charge distribution, and $s_1(N_A)= P(N_A)$ is the probability of having $N_A$ particles in region $A$, which has been calculated before~\cite{song2012bipartite}. 
Eq.~(\ref{explicitsn}) with $n=1$ implies that the variance in the number of particles in a segment of length $L$ is $\langle \Delta N_A^2 \rangle = \frac{K \ln L}{\pi^2}$
\footnote{A relation between $P(N_A)$ and the entanglement entropy $\mathcal{S}(L)$ was recently suggested~\cite{klich2009entnoise,song2012bipartite}.}.

The charge-resolved EE can now be calculated from $\mathcal{S}(N_A) =- \partial_n s_n(N_A)|_{n \to 1}$, giving
\be
\label{eq:approx}
\mathcal{S}(N_A) =\frac{1}{3}\sqrt{\frac{\pi \ln L }{2K}}  e^{- \frac{ \pi^2 \Delta N_A^2}{2 K \ln L} }  - \mathcal{O} \left( \frac{1}{\sqrt{\ln L }} \right).
\ee
We can see that the decomposition of the total entanglement $\mathcal{S}(L) = \frac{c}{3} \ln L$ into charge contributions is controlled by $P(N_A)$, the Gaussian distribution of charge in region $A$. The maximal contribution $\mathcal{S}(\langle N_A \rangle)$ scales as $\sqrt{\ln L}$, which  is plausible given the $\sqrt{\ln L}$ scaling of the standard deviations of charge fluctuations.
Eq.~(\ref{eq:approx}) was recently conjectured based on numerical data~\cite{laflorencie2014}.

We checked our CFT predictions via numerical simulations for noninteracting ($K=1$) fermions on a lattice.
In general one may define the entanglement Hamiltonian $H_A$ by $\rho_A = e^{-H_A}$. For a noninteracting system 
$H_A$ is quadratic in the fermionic operators.
Denoting its single-particle eigenvalues by $\varepsilon_l$, the expressions for the entropies are similar to the thermal entropies of free fermions with unit temperature and Hamiltonian $H_A$,
\modified{%
\be
  s_n(\alpha) = \prod_l \left[ e^{i\alpha} (f_l)^n + (1-f_l)^n \right]
\ee
where $f_l = 1/(e^{\varepsilon_l}+1)$ can be easily obtained numerically as the eigenvalues of the equal-time two-point fermionic correlations matrix, $C_{ij} = \langle c^\dagger_i c_j \rangle$ ($i,j=1\cdots L$)~\cite{peschel2003}.}
Choosing a subsystem of $L=10000$ sites of an infinite half-filled tight-binding chain (for which $C_{ij} = \sin [\pi (i-j)/2]/[\pi(i-j)]$) we computed the distribution of occupancies $P(N_A) = s_1(N_A)$, and the particle-number resolved entanglement $\mathcal{S}(N_A)$. Fig.~2 shows the numerical results as dots and our analytical formula with $K=1$ without any fitting parameter (but including $\mathcal{O}( (\ln L)^0 )$ corrections~\cite{jin2004,song2012bipartite}) as continuous lines. As also seen in Fig.~2, even for a large subsystem of $L=10000$, $\ln L$ is moderately large and the distribution is quite narrow. Thus, in practice it is inaccurate to keep only the leading contribution in $\sqrt{\ln L}$ in Eq.~(\ref{eq:approx}) and instead one has to evaluate Eq.~(\ref{eq:saNA}) with Eq.~(\ref{mainresult}).

\begin{figure} 
	\centering	
	\includegraphics[scale=0.35]{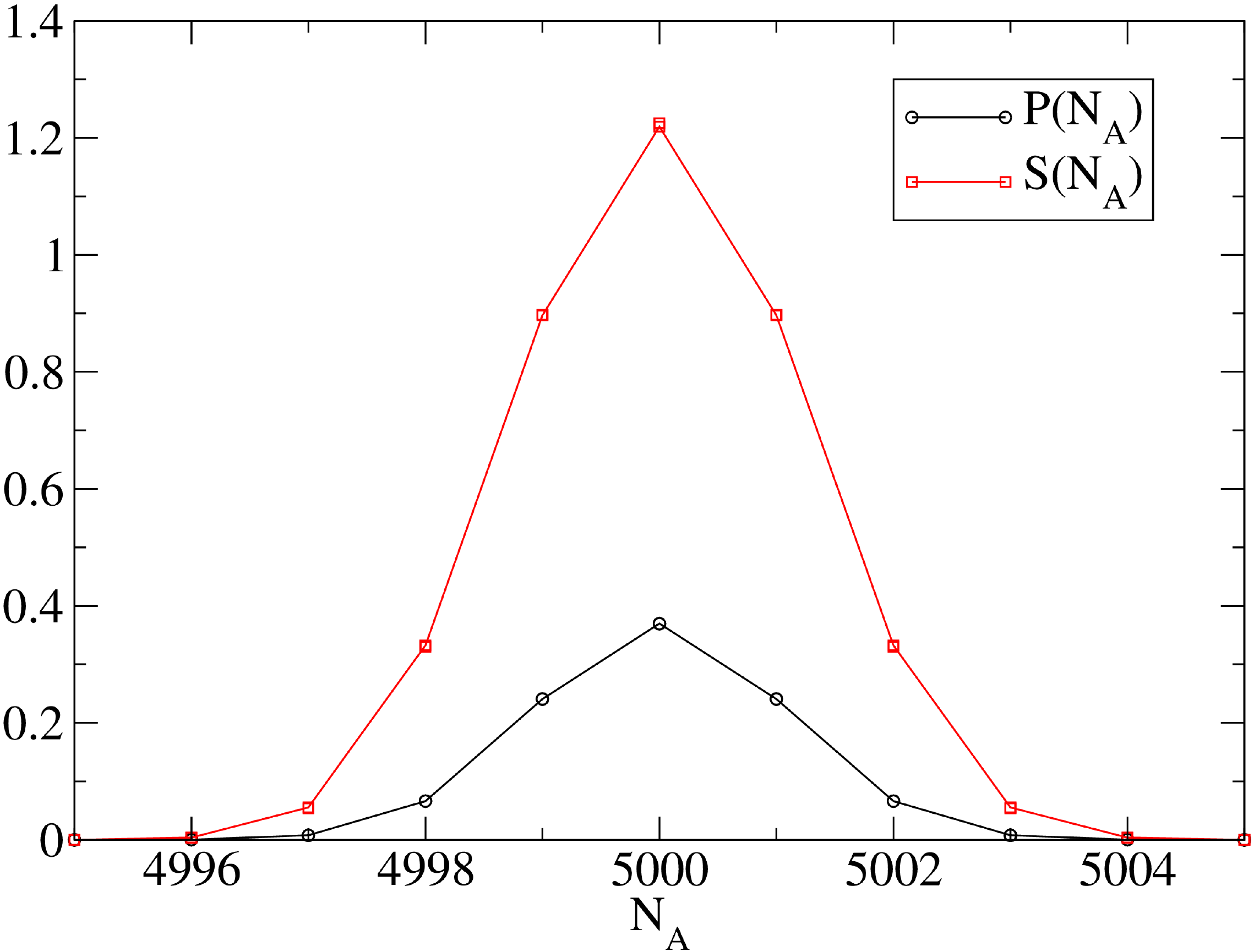}
	\caption{(Color online) Charge distribution $P(N_A)$ and charge-sector contributions to entanglement entropy $\mathcal{S}(N_A)$ in  a subsystem of $L=10000$ sites of an infinite half-filled tight-binding chain, computed numerically (dots) and analytically (continuous lines).
	}
	\label{fig:2}
\end{figure}

The analytic dependence of Eq.~(\ref{mainresult}) on $n$ can be used to extract further information including the full entanglement spectrum of each charge block of the density matrix, $\{\lambda_i(N_A) \}$ (so that $s_n(N_A) = \sum_i [\lambda_i(N_A)]^n $). 
Let us first consider the maximal eigenvalue, $\lambda_{\mathrm{max}}(N_A) = \lim_{n\to\infty} [s_n(N_A)]^{1/n}$.
Using Eq.~(\ref{mainresult}) we find
\be
\label{lambdamax}
-\ln \lambda_{\mathrm{max}}(N_A) = \frac{1}{6} \ln L +  \frac{ \pi^2 }{2 K \ln L} \Delta N_A^2.
\ee
While the first term was derived by Calabrese and Lefevre~\cite{calabrese2008entanglement} we here obtain the dependence on the particle number $N_A$
\footnote{The $\Delta N_A=0$ part also has $1/\ln L$ corrections, which require a more careful evaluation of $s_n$~\cite{orus2006lambdamax,calabrese2010lambdamax}.}.
Let us recall that $- \ln \lambda_i$ are the many-body eigenvalues of the entanglement Hamiltonian $H_A$.
For non-interacting fermions, 
the low-lying single-particle eigenstates of $H_A$
have been calculated analytically~\cite{peschel2009reduced}, $\varepsilon_l = \pm \frac{\pi^2}{2 \ln L} (2l-1)$, $l=1,2,3...$. Filling up the negative energy states ($\Delta N_A$=0) and then  adding $\Delta N_A$ particles 
gives $\frac{\pi^2}{2 \ln L} \sum_{l=1}^{\Delta N_A} (2l-1) = \frac{\pi^2}{2 \ln L} \Delta N_A^2$, in exact agreement with our general result Eq.~(\ref{lambdamax}).
Our field-theory approach generalizes these results to the interacting case, $K \ne 1$, confirming a previous conjecture based on numerics~\cite{laflorencie2014}.

Turning to the full entanglement spectrum, let us denote its density for a given charge sector by $P(\lambda, N_A)$. Defining the integrated density, $n(\lambda, N_A) = \int_{\lambda}^{\lambda_\mathrm{max}(N_A)} d\lambda^\prime P(\lambda^\prime,N_A) $, and
using the methods of Calabrese and Lefevre~\cite{calabrese2008entanglement}, we find (see the Supplemental Material for details~\cite{Note2})
\be
\label{eqn:spectrum}
n(\lambda, N_A) = \int_{0}^{\pi} \frac{d\alpha}{\pi} \cos(\alpha N_A) I_0 \left( 2 \sqrt{ r(\alpha) \ln \frac{1}{\lambda_\mathrm{max}} \ln \frac{\lambda_\mathrm{max}}{\lambda} } \right),
\ee
where $\lambda_\mathrm{max} \equiv \lambda_\mathrm{max}(\Delta N_A=0)$, $r(\alpha) = 1- 3 K \alpha^2/\pi^2$, and $I_0(z)$ is a modified Bessel function~\cite{gradshteyn}.
The same quantity can be calculated numerically for the noninteracting ($K=1$) tight-binding chain, and the results nicely agree with the CFT prediction, as can be seen in Fig.~3.

\begin{figure} 
	\centering	
	\includegraphics[scale=0.35]{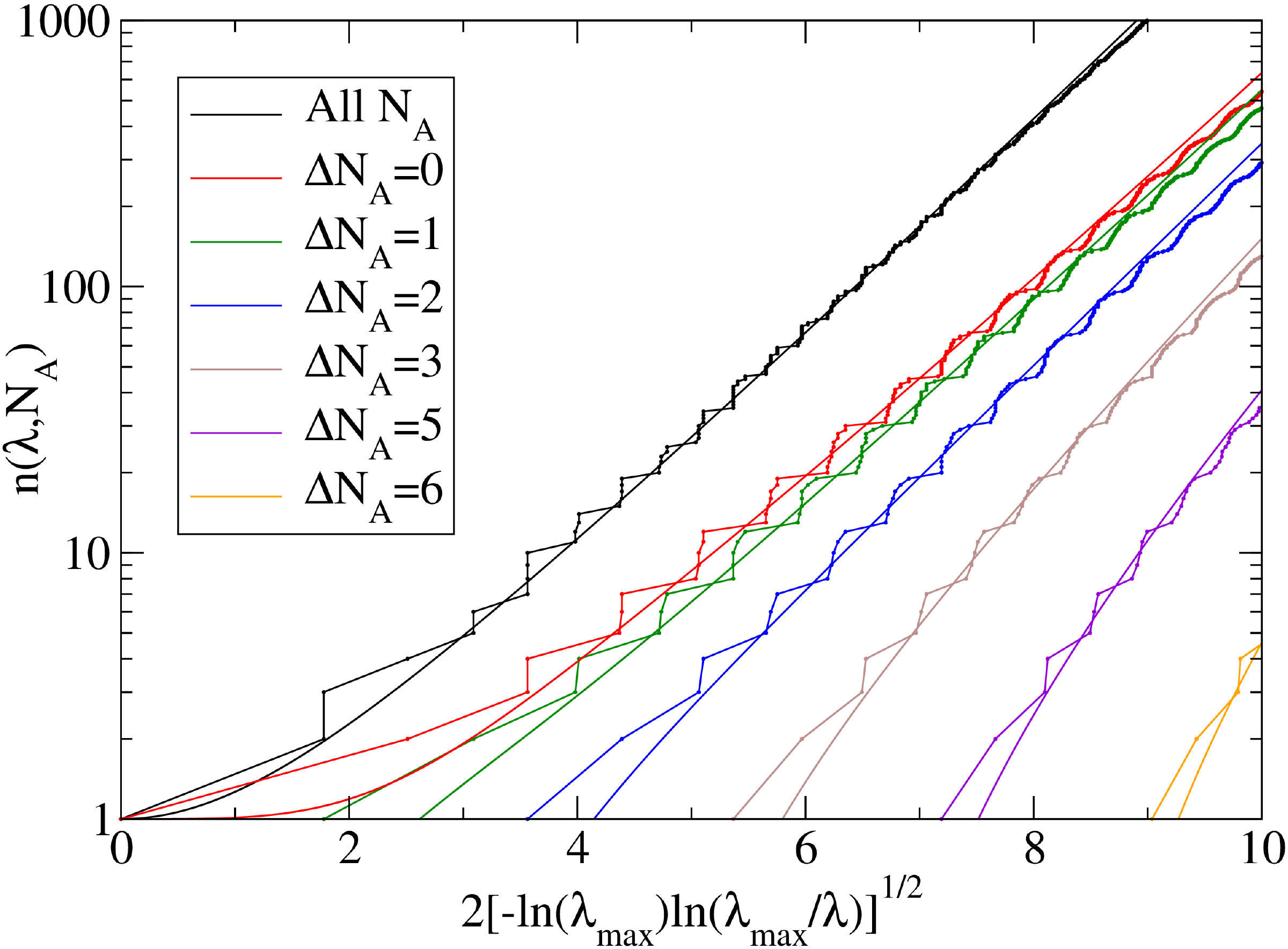}
	\caption{(Color online) Integrated density of the entanglement spectrum $n(\lambda, N_A)$ for the same system in Fig.~2 computed numerically (discontinuous lines) and analytically (continuous lines) for various particle numbers $N_A$. Numerically we used the 24 closest-to-zero single particle eigenvalues of the entanglement Hamiltonian to build the highest many-body eigenvalues of $\rho_A$.
	}
	\label{fig:3}
\end{figure}

\emph{SU(2) symmetry.---}
The possibility to decompose entanglement measures into charge sectors via the insertion of an \modified{Aharonov-Bohm} flux in the Riemann geometry applies to general symmetries. 
Now we will demonstrate this for a nonabelian symmetry, and consider $SU(2)$ spin chains as a case study. Our basic goal is now to break the entropy of an $SU(2)$ symmetric system into contributions with a fixed total spin $\vec{S}_A^2$ as well as total projection $S^z_A$ in region $A$.
In order to compute $\mathrm{Tr}_{S_A} \rho_A^n$ for a fixed spin $S_A$ of region $A$, $\vec{S}_A^2= S_A(S_A+1)$, one can use an identity valid 
for any spin-rotation-symmetric operator such as $\rho_A^n$
\footnote{Which is the case if the overall system is in an $SU(2)$ invariant state. If the overall system spin is nonzero (e.g., a ferromagnet), this requires building an equally-mixed state with all possible $S^z$ values of the overall system.}, 
\be
\label{trick}
{\rm{Tr}}_{S_A,S_A^z} \rho_A^n=\mathrm{Tr}_{S_A^z = S_A} \rho_A^n - \mathrm{Tr}_{S_A^z = S_A+1} \rho_A^n. 
\ee
The right hand side involves the quantity $s_n(S^z_A) \equiv \mathrm{Tr}_{S^z_A} \rho_A^n$, which is a sum over states with fixed $S^z_A$ in region $A$. The latter can be computed along the same methods developed for the $U(1)$ case above. 

We will now calculate $s_n(S^z_A)$ for critical spin chains. A family of critical $SU(2)$ symmetric theories are the $SU(2)_k$ Wess-Zumino-Witten (WZW) models~\cite{affleck1987critical}. For $k=1$ they describe the familiar spin-1/2 Heisenberg chain, while for other integer $k$ they correspond to certain critical spin-$k/2$ chains~\cite{affleck1987critical}.
Let us first recall that in the $U(1)$ case, the operator $e^{i \alpha \hat{N}_A}$ appearing in Eq.~(\ref{eq:definesn}) can be written as $e^{i \frac{\alpha}{2 \pi} \int_A dx \partial_x \phi}$, which indeed becomes a product of the vertex operators mentioned above, $\mathcal{V}(w) \mathcal{V}^\dagger(w^\prime) = e^{i \frac{\alpha}{2 \pi}\phi(w) } e^{-i \frac{\alpha}{2 \pi}\phi(w')}$ (here we write only holomorphic factors). Similarly, in a spin chain we have the operator $e^{i \alpha \hat{S}^z_A} = e^{i \alpha \int_A dx J^z}$, with $\vec{J}(z)$ the WZW spin current. 
The corresponding vertex operators have the scaling dimensions
\be
\Delta_\mathcal{V}^{(k)}=			\bar{\Delta}_\mathcal{V}^{(k)} =\frac{k}{4}\left( \frac{\alpha}{2\pi} \right)^2, ~~~k \in \mathbb{N}.
\ee
As a check, for $k=1$ we have the Heisenberg chain, which is equivalent (via the Jordan-Wigner transformation) to interacting spinless fermions with $K=1/2$~\cite{gogolin}, in agreement with our previous $U(1)$ results.
Using Eqs.~(\ref{eq:saNA}) and (\ref{trick}) we obtain (for $\ln L \gg 1$)
\be
\mathcal{S}(S_A,S^z_A) = (2S_A+1) \frac{c \pi^{5/2}}{3 k^{3/2} \sqrt{\ln L}} e^{-\frac{\pi^2 S_A^2}{k \ln L}},
\ee
where $c = \frac{3k}{k+2}$ is the central charge of the WZW model. This equation displays a further reduction of the scaling of the entropy; since the typical value of $S_A$ scales as $\sqrt{\ln L}$, $\mathcal{S}(S_A,S^z_A)$ scales as 
$\mathcal{O} (L^0)$.

\emph{Discrete symmetries.---}
To demonstrate the method for discrete symmetries, consider a $\mathbb{Z}_N$ charge $Q \mod N$ with $Q = Q_A+Q_B$. The system under consideration can be the clock model, or a chain of parafermions~\cite{zamolodchikov1985nonlocal,alicea2016parafermion}. For a $Z_N$ symmetric state we can decompose the entropies according to the subsystem charge, $s_n(Q_A) = \frac{1}{N}\sum_{\alpha=0}^{N-1}e^{-i \frac{2 \pi \alpha}{N} Q_A} s_n(\alpha)$, where $s_n(\alpha) = \mathrm{Tr} \rho_A e^{i \frac{2 \pi \alpha}{N} Q_A}$ $(\alpha,Q_A = 0,1,...,N-1)$. Does $s_n(Q_A)$ actually depend on $Q_A$ and how? 

As a transparent example, consider $N=2$, and specifically the quantum Ising chain $H=-J \sum_i \sigma^z_i  \sigma^z_{i+1} - h \sum_i \sigma^x_i$, \modified{which is equivalent to a chain of Majorana fermions via the Jordan-Wigner transformation~\cite{kitaev2001}. 
Here the components of the total spin (fermion number) are not conserved, but the parity $(-1)^Q=\prod_i \sigma_i^x$ of the number of spins in the $+x$ direction (fermion number parity) is, so entanglement can be decomposed into the two sectors of even/odd $Q_A$.}
Using the duality transformation to disorder fields which are new Pauli operators $\mu^z_i = \prod_{j \le i} \sigma_i^x$, $\mu_i^x = \sigma^z_i  \sigma^z_{i+1}$ (in terms of which the Hamiltonian attains the same form but with $J \leftrightarrow h$), we express the desired counting operator as $(-1)^{Q_A}=\prod_{j \in A} \sigma^x_j = \mu_{1}^z \mu_{L}^z$, where region $A$ extends from site 1 to $L$. 
Moving to the critical state at $J=h$, described by a $c=1/2$ Ising CFT, the disorder operator $\mu$ has scaling dimension $\Delta_\mu =\bar{\Delta}_\mu =\frac{1}{16}$. 
Plugging this into the above results (with $\mathcal{V} = \mu$) we get
\be
s_n(Q_A) =L^{-(n-1/n)/12} \frac{1}{2} (1  +  L^{-1/(4n)} (-1)^{Q_A}),
\ee
a result we have verified numerically.
\modified{Setting $n=1$ one obtains $P(Q_A)$, the probability of finding a given parity 
in region $A$.  
As expected, the dependence on $Q_A$ disappears at $L \to \infty$.} One can readily generalize the calculation to $Z_N$ models such as the clock model or parafermions, using the parafermion CFT~\cite{zamolodchikov1985nonlocal} 
where the central charge is $c=\frac{2(N-1)}{N+2}$ and the scaling dimension of the generalized disorder operators is $\Delta_{\mu_\alpha} = \bar{\Delta}_{\mu_\alpha} =\frac{\alpha(N-\alpha)}{2N(N+2)}$, ($\alpha=0,...,N-1$).

\emph{Experimental measurement.---}
While so far we treated the replica construction as a purely theoretically trick, in a remarkable recent experiment~\cite{islam2015measuring} it has been applied 
in the lab, demonstrating for the first time the possibility to perform a measurement of entanglement in a many-body system, specifically the second RE. Following a theoretical prediction~\cite{daley2012measuring}, their protocol for measuring $s_2$ consists of (i) preparing 2 copies using optical lattice techniques, (ii) performing a transformation between the copies using a  Hong-Ou-Mandel interference, and (iii) a parity measurement of the charge in region $A$ in a specific copy. One can then easily modify the last stage by measuring the charge of region $A$ in both copies and calculating the average $N_A$ \modified{(since only integer values of this average contribute~\cite{daley2012measuring,cornfeld18})}. The average parity of one copy for given $N_A$ would yield $s_2(N_A)$.
\modified{One may also multiply the parity by $e^{i\alpha N_A}$ and average over all $N_A$ to experimentally obtain the ``flux RE'' $s_2(\alpha)$.}
The extension to $n>2$ is similar.
Let us note that a recent work brought up another route for the experimental measurement of the RE without using replicas, which actually gives access to the charge-resolved entropies as well~\cite{elben2017measuring}.

\emph{Future outlook.---}
Many interesting questions arise from our results, including: the scaling of the charge-resolved entanglement in higher dimensions or in the presence of boundary critical phenomena~\cite{affleck2009entanglement,PhysRevA.74.050305,PhysRevB.96.075153}, its behavior in topological systems, and other entanglement measures such as the negativity~\cite{calabrese2012negativity,cornfeld18} and the relative entropy~\cite{lashkari2014relative}.

\begin{acknowledgments}
\emph{Acknowledgements.---}
We thank Y. Avron for posing a question that triggered this project and E. Bettelheim for insightful key remarks in the initial stages of this work.
M.G. was supported by the Israel Science Foundation (Grant No. 227/15),
the German Israeli Foundation (Grant No. I-1259-303.10), the US-Israel Binational Science Foundation (Grant No. 2014262), and the Israel Ministry of Science and
Technology (Contract No. 3-12419).
E.S. was supported by the Israel Science Foundation (Grant No. 1243/13), and by the  the US-Israel Binational Science Foundation (Grant No. 2016255).
\end{acknowledgments}

\bibliographystyle{apsrev4-1}
\bibliography{chargedeeref}	


\renewcommand{\thesection}{S.\Alph{section}}
\setcounter{figure}{0}
\renewcommand{\thefigure}{S\arabic{figure}}
\setcounter{equation}{0}
\renewcommand{\theequation}{S\arabic{equation}}

\newpage\pagebreak

\section*{Supplemental Material for ``Symmetry-resolved entanglement in many-body systems''}

In this Supplemental Material we present some technical details which were omitted in the main text.
In Sec.~\ref{appendix1} we derive the scaling dimension of the twist field $\mathcal{T}_\mathcal{V}$, Eq.~(4) of the main text.
In Sec.~\ref{appendix2} we calculate the density of entanglement eigenvalues resolved by charge, Eq.~(11) of the main text.



\section{Scaling dimension of the twist field $\mathcal{T}_\mathcal{V}$} 
\label{appendix1}
In this section we apply the methods of Ref.~[5] in order to extract the scaling dimension of our composite twist field $\mathcal{T}_\mathcal{V} = \mathcal{V}\mathcal{T}$ [Eq.~(4) of the main text], where $\mathcal{V}$ generates the Aharonov-Bohm flux and $\mathcal{T}$ generates the Riemann geometry. Using the relation, Eq.~(3), for correlation functions on $\mathcal{R}_n$, we consider the case where $\mathcal{O}(z)$ is the total stress-energy tensor $T(z)=\sum_{l=1}^n T_l(z)$. Its expectation value in $\mathcal{R}_n$ with flux $\alpha$ can be evaluated using the uniformizing mapping $\xi(z) = \left( \frac{z-w}{z-w'} \right)^{1/n}$ from $\mathcal{R}_n$ to the complex plane $\mathcal{C}$ with flux $\alpha$,
\be
\label{expT}
\langle T(z) \rangle_{\mathcal{R}_n} =\sum_l \left(\frac{d\xi}{dz} \right)^2 \langle T_l \rangle_{\mathcal{C},\alpha} + \frac{c n}{12} \{\xi,z \},
\ee
where the Schwarzian derivative $\{\xi,z \}$ is given by
\bea
\label{schw}
\frac{c}{12} \{\xi,z \} = \frac{c(1-n^{-2})}{24} \frac{(w'-w)^2}{(z-w)^2(z-w')^2} .
\eea
The conformal transformation $z \to \xi(z)$ takes $(w,w')$, the end points of region $A$, to $(0,\infty)$, and subsequently takes an $n$th root. Hence it converts a closed orbit in $\mathcal{R}_n$ circling $w$ $n$ times into a single-winding orbit around $\xi(w)=0$. The first term in the transformation Eq.~(\ref{expT}) contains $\langle T_l \rangle_{\mathcal{C},\alpha}$, the expectation value of stress-energy tensor in the new coordinates on the single plane. In the absence of the Aharonov-Bohm flux it vanishes~[5]. However, this coordinate transformation does not remove the flux which now pierces the plane at $\xi(w)=0$ and $\xi(w')=\infty$ in opposite directions. Thus, winding a particle around the origin or infinity still leads a phase, and 
\be
\langle T_l \rangle_{\mathcal{C},\alpha}=		\frac{\langle T_l \mathcal{V}(\xi(w)) \mathcal{V}(\xi(w')) \rangle_{\mathcal{C}} }{\langle \mathcal{V}(\xi(w)) \mathcal{V}(\xi(w')) \rangle_{\mathcal{C}}}.
\ee
As in Eq.~(\ref{expT}), the denominator ensures that the expectation value of the identity operator gives unity; The transformation factors $(dw/dz)$ of the scaling field $\mathcal{V}$ cancel between the numerator and denominator.
Assuming that $\mathcal{V}$ is a primary field of (holomorphic) dimension $\Delta_\mathcal{V}$, we have
\be
\langle T_l \rangle_{\mathcal{C},\alpha}=\Delta_\mathcal{V}\frac{(\xi(w)-\xi(w'))^2}{(\xi(z)-\xi(w))^2 (\xi(z)-\xi(w'))^2} \to \frac{\Delta_\mathcal{V}}{\xi(z)^2}.
\ee
In the last relation we took the limit $\xi(w) \to 0$, $\xi(w') \to \infty$.  
Putting everything together, we see that 
the composite twist field behaves as a primary field with scaling dimension $\Delta_n(\alpha)$,
\be
\frac{\langle T(z)  \mathcal{T}_\mathcal{V}(w) \mathcal{T}_\mathcal{V}(w') \rangle_{\mathcal{C}^n} }{\langle \mathcal{T}_\mathcal{V}(w) \mathcal{T}_\mathcal{V}(w') \rangle_{\mathcal{C}^n}} =\Delta_n(\alpha) \frac{(w-w')^2}{(z-w)^2 (z-w')^2},
\ee
with $\Delta_n(\alpha)=\frac{ c(n-n^{-1})}{24}+\frac{\Delta_\mathcal{V}}{n}$, hence proving Eq.~(4). We note in passing that the $1/n$ factor in the last term resembles the scaling of corrections to the entropies~[45], and that similar techniques may be employed for calculating entanglement in excited states created by primary operators~[46].

\section{The charge-resolved entanglement eigenvalue density} 
\label{appendix2}
Starting from Eq.~(5) of the main text, we will now derive the integrated entanglement eigenvalue density, Eq.~(11) of the main text, generalizing the approach of Ref.~[27] to the charge resolved case.
Let us define $P(\lambda,\alpha) = \sum_{N_A} P(\lambda,N_A) e^{i \alpha N_A}$.
Then  $\lambda P(\lambda,\alpha) = \mathrm{Im} f(\lambda-i0^+,\alpha)$, where
\be
f(z,\alpha) = \frac{1}{\pi} \sum_{n=1}^{\infty} s_n(\alpha) z^{-n} = \frac{1}{\pi} \int d\lambda \frac{\lambda P(\lambda,\alpha)}{z-\lambda}.
\ee
By Eq.~(5), $s_n(\alpha) = c_n(\alpha) e^{-b [ n -r(\alpha)/n]}$ (for $c=1$), with $b=-\ln\lambda_\mathrm{max}=\frac{1}{6}\ln L$, and $r(\alpha) = 1- 3\alpha^2 K /\pi^2$. Assuming that $c_n(\alpha)$ is close to unity (in practice, only the exact behavior of $P(\lambda,N_A)$ very close to $\lambda_\mathrm{max}(N_A)$ is sensitive to it) we have
\begin{align}
f(z,\alpha) = & \frac{1}{\pi} \sum_{k=0}^\infty \frac{(b r(\alpha))^k}{k!} \sum_{n=1}^\infty \frac{1}{n^k} \left( \frac{\lambda_\mathrm{max}}{z}  \right)^n
\nonumber \\
= &
\frac{1}{\pi} \sum_{k=0}^\infty \frac{(b r(\alpha))^k}{k!} \mathrm{Li}_k(\lambda_\mathrm{max}/z),
\end{align}
where $\mathrm{Li}_k(y)$ is the polylogarithm function, which obeys
$\mathrm{Im} [ \mathrm{Li}_k(y+i0^+) ] = \pi \theta(y-1)(\ln y)^{k-1}/\Gamma(k)$
for $k\ge 1$, with $\theta(x)$ and $\Gamma(x)$ the step and Gamma functions, respectively~[30]. Integrating over $\lambda$ and summing over $k$ we find
\be
n(\lambda,\alpha) \equiv \int_\lambda^{\lambda_\mathrm{max}(N_A)} d\lambda^\prime P(\lambda^\prime,\alpha) = I_0 \left( 2 \sqrt{br(\alpha) \ln \frac{\lambda_\mathrm{max}}{\lambda} } \right),
\ee
with $I_0(z)$ the modified Bessel function, from which Eq.~(11) immediately follows. 

\end{document}